\documentclass[%
 reprint,superscriptaddress,
 amsmath,amssymb, nofootinbib,
 aps,twocolumn,pra
]{revtex4-1}
\usepackage{graphicx}
\usepackage{appendix}
\usepackage{multirow}
\usepackage{amsfonts}
\usepackage{float}
\usepackage{bbold} 
\usepackage{verbatim} 
\usepackage{ulem} 
\usepackage[width=.8\textwidth]{caption}
\usepackage[margin=0.01cm]{caption}
\usepackage[usenames, dvipsnames]{xcolor}
\usepackage{xifthen}

\newcommand{\ket}[1]{| #1 \rangle}

\newcommand{\arxiv}[2][]{\ifthenelse{\isempty{#1}}{\href{http://arxiv.org/abs/#2}{{\tt arXiv:\allowbreak{}#2}}} {\href{http://arxiv.org/abs/#2}{{\tt arXiv:\allowbreak{}#2 [#1]}}}}

\begin{document}

\title{Delegated verification of quantum randomness with linear optics}

\author{Rodrigo Piera\smallskip}
\affiliation{Quantum Research Centre, Technology Innovation Institute, Abu Dhabi, United Arab Emirates}
\author{Jaideep Singh\smallskip}
\affiliation{Quantum Research Centre, Technology Innovation Institute, Abu Dhabi, United Arab Emirates}
\author{Yury Kurochkin\smallskip}
\affiliation{Quantum Research Centre, Technology Innovation Institute, Abu Dhabi, United Arab Emirates}
\author{James A. Grieve\smallskip}
\affiliation{Quantum Research Centre, Technology Innovation Institute, Abu Dhabi, United Arab Emirates}





\begin{abstract}
Randomness is a critical resource of modern cryptosystems. Quantum mechanics offers the best properties of an entropy source in terms of unpredictability. However, these sources are often fragile and can fail silently. Therefore, statistical tests on their outputs should be performed continuously.
Testing a sequence for randomness can be very resource-intensive, especially for longer sequences, and transferring this to other systems can put the secrecy at risk. In this paper, we present a method that allows a third party to publicly perform statistical testing without compromising the confidentiality of the random bits by connecting the quality of a public sequence to the private sequence generated using a quantum process. We implemented our protocol over two different optical systems and compared them.
\end{abstract}

\maketitle

\section{Introduction}

Quantum Mechanics revolutionizes our understanding of randomness, presenting a paradigm shift from the deterministic frameworks of classical physics \cite{khrennikov2016probability,volovich2011randomness,coecke2008classical}. Unlike classical processes, which operate under the assumption of predictability given sufficient information, quantum phenomena introduce a layer of randomness that is intrinsic and unavoidable. This quantum randomness is not merely a result of observational limitations but a fundamental characteristic of the quantum realm, offering a novel perspective on the nature of unpredictability
\cite{bera2017randomness,leggett2002testing}. Inside the Quantum mechanics framework, this unpredictability takes the form of the superposition principle, in which a quantum object's wave function can exist simultaneously in several states until it is measured. Upon measurement, the superposed state collapses into one of the possible outcomes, leading to an inherently random result \cite{yuan2015intrinsic,busch2016quantum}.

This randomness, beyond its theoretical interest, has practical applications, notably in enhancing the security of cryptographic systems, whether classical or quantum-based \cite{pirandola2020advances,ma2016quantum,jacak2021quantum}. Emerging from these principles, Quantum Random Number Generators (QRNGs) (\cite{jennewein2000fast,sanguinetti2014quantum,singh2024compact}) harness quantum processes to generate unpredictable numbers, offering an advanced level of security and reliability in various applications \cite{ma2016quantum,herrero2017quantum}.
The ideal functioning of a QRNG can be summarised as follows:

\begin{enumerate}
    \item Preparation of an input state in a superposition of measurable states, for example $\frac{\ket{0}+\ket{1}}{\sqrt{2}}$ and
    \item Measurement of the state from which we obtain the random bits (zero or one, depending on the output).
\end{enumerate}

However, the practical implementation of a QRNG is more complex due to the imperfection of real devices \cite{frauchiger2013true}. Environmental influences such as temperature fluctuations or electromagnetic interference can lead to distortions in the preparation of the state and measurement, affecting the random sequence. Typically, these influences that the manufacturer of the devices does not fully control are called noise. One solution to overcome these problems is using post-processing, which requires modeling the noise source to determine how many epsilon secure bits we can extract from the random process \cite{ma2013postprocessing}. However, for this technique, it is assumed that the noise source and internal parameters remain constant until they are rechecked. But what happens if some internal element stops working correctly without being detected? A Random number generator that is not working properly could cause security problems, mainly when used in cryptosystems. 
The physical random number generator industry has several tests to monitor potential randomness issues, many of them based on the analysis of the generated sequence. However, the complexity of the test often requires a significant amount of computing power and time, especially for longer sequences. The sequence size required for a statistically valuable test for fluctuation and correlation can also be quite large \cite{jacak2020quantum}, about $10^9$ bits.

Resource-constrained environments such as Internet of Things devices or embedded systems may not have the processing power, memory, or energy resources required for extensive random testing. Meanwhile, according to NIST and other industry standards, these devices must be protected against quantum attacks with post-quantum cryptography \cite{nist}. Post-quantum cryptography relies on the quality of randomness, making it essential to have a reliable QRNG device for these applications.



To address this issue, \cite{jacak2020quantum} proposed a theoretical protocol that allows transferring the testing task to a trusted third party without violating the secrecy of the private bit sequence generated locally into the device. This protocol requires generating a tripartite entangled state to perform the task and is highly susceptible to detector efficiency drift or bias in the state preparation or measurement stages. The authors' proposed solution to these problems requires the consumption of random numbers. This protocol was adapted in \cite{islam2024privacy} to work with a pair of entangled photons instead of a tripartite entangled system.

In this paper, we present a simplified method that allows third-party verification of a sequence generated by a QRNG without relying on entangled pairs generation. 
Our analysis is based on the hypothesis that we are operating a trusted device that is not ideal, and its malfunctioning could lead to errors in the output or bit imbalance due to variations in the efficiency of the detectors. Therefore, before using the random sequence generated by our QRNG, it must pass a series of public tests performed outside the device without using local resources. If this sequence passes these tests, the private sequence can be post-processed (if necessary) to eliminate any information that may have been leaked by the device or classical noise in the system, and the final sequence can be used. We provide a proof-of-concept experiment and discuss strategies for reducing potential sources of error.



\section{PROTOCOL DESCRIPTION}


Our protocol is based on the generation of two random sequences, $Q_1$ and $Q_2$, ensuring they share the same statistical properties (i.e., $H(Q_1)=H(Q_2)$) while maintaining zero mutual information (i.e., $I(Q_1,Q_2)=0$). The sequence $Q_2$ (public sequence) is sent to an external party, which tests it (using, for example, the Dieharder test) to verify if $Q_2$ meets the characteristics of a random sequence. Since both sequences have the same entropy, verifying the statistical properties of $Q_2$ indirectly verifies $Q_1$. With zero mutual information between the sequences, there is no information leakage.

To generate sequences with these properties, we prepare a four-level superposition state:
\begin{equation}
\ket{\Phi}=\big(\sqrt{p_{1}}\ket{00}+\sqrt{p_{2}}\ket{01}+\sqrt{p_{3}}\ket{10}+\sqrt{p_{4}}\ket{11}\big).
\label{eq:state}
\end{equation}
\begin{figure}
\centering
\includegraphics[scale=0.3]{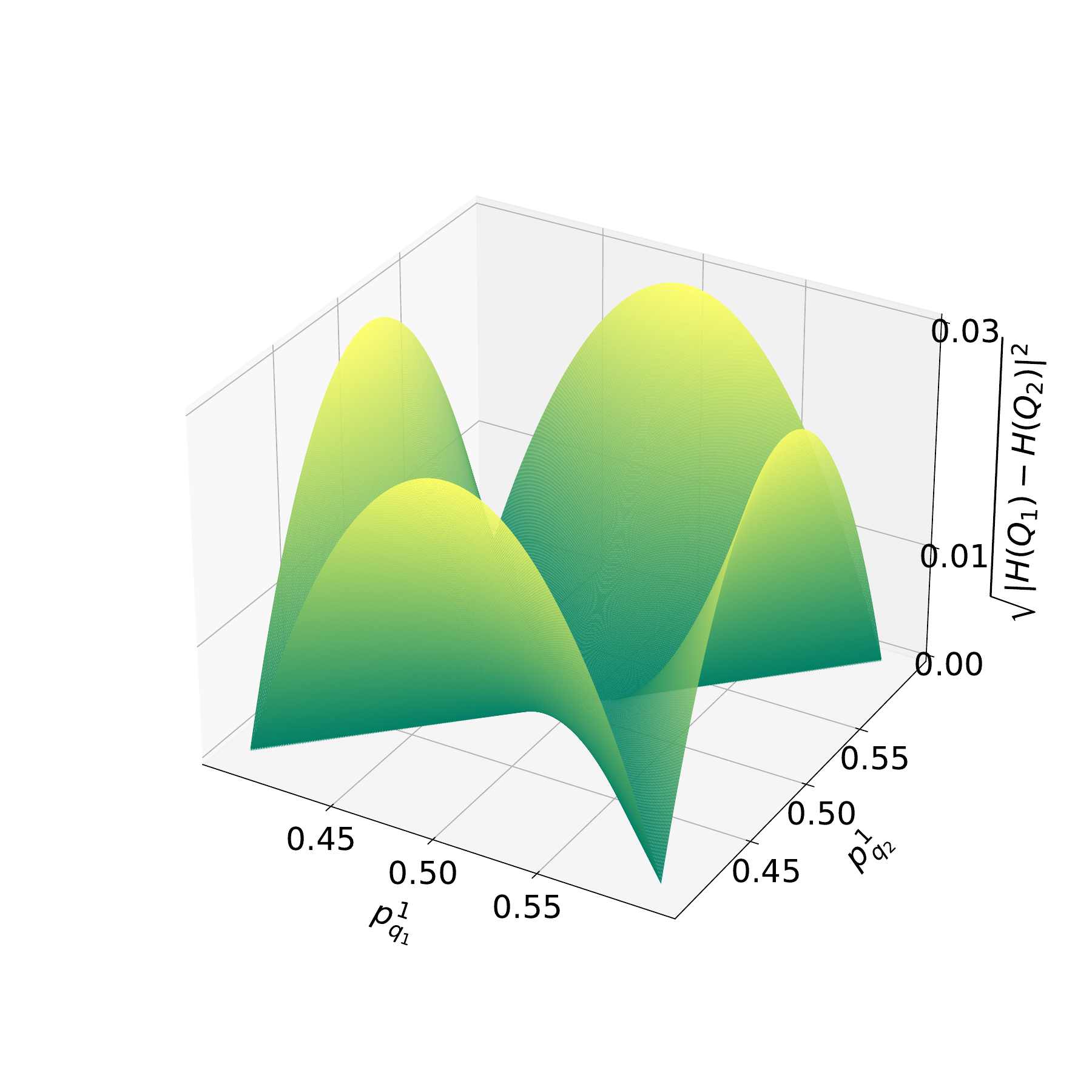}
  \caption{Difference in entropy between $Q_{1}$ and $Q_{2}$ as a function of $p_{q_{1}}^{1}$ and $p_{q_{2}}^{1}$. Allowing fluctuations around 0.1 results in a maximum distance of $0.028$.}
  \label{fig:Entropy_diff}
\end{figure}
In this equation, the bit pairs inside the kets represent the codification of bits in sequences $Q_1$ and $Q_2$. For instance, if the state $\ket{01}$ is measured, we assign bit $0$ to $Q_1$ and bit $1$ to $Q_2$. This state is measured $N$ times, producing two sequences with $N$ bits each. If $pi=0.25$ for all i, $Q_1$ and $Q_2$ meet the required conditions for public verification.



In practical applications, system instability and noise can cause protocol failure. Therefore, it is necessary to account for state preparation and measurement fluctuations, correlations between outputs, and system noise. Fluctuations in state preparation or measurement can impact the protocol by causing a drift in probabilities ($dp_{i}/dt\neq 0$) or introducing bias ($dp_{i}/dt= 0$ but $p_i\neq p_j$), leading to incorrect statistical equivalence between the sequences. Correlations between outputs can cause similar errors and information leakage from the private sequence via the public one.

For noise, the issue is different. If the noise-generating process is classical, a classical model might predict some bits in the sequence, weakening its unpredictability, which is highly valued in QRNG. Since noise cannot be completely eliminated, random amplification routines are proposed \cite{frauchiger2013true,ma2013postprocessing,bennett1988privacy}. These routines map a sequence of bits to a shorter one, resulting in a sequence with uniformly distributed properties, ensuring that partial knowledge of the raw sequence does not reveal information about the shorter one. 

Not all noise sources are classical, but for security, it is prudent to consider any external randomness contribution as classical noise, even if it might have a quantum mechanical origin.

To mitigate these issues, it is possible to characterize the drift's time scale and take measurements at intervals where the process remains stable and unbiased. This involves ensuring measurements are taken over periods shorter than the drift of the process and checking the output frequency. The sequence $Q_2$ is sent for public verification when the output frequencies fall within a predefined interval. For example, in Fig. \ref{fig:Entropy_diff}, we show the entropy difference between the two random sequences generated by state \ref{eq:state} as a function of the frequency of the element one in both sequences ($p_{Q_1}^{1}=p_3+p_4$ and $p_{Q_2}^{1}=p_2+p_4$).

It is also important to check if the detector outputs are independent and memoryless. Although our protocol involves local calculations, none require resource-intensive tasks.



Finally, we can sum up our protocol in the following way:

\begin{enumerate}

    \item Prepare and measure the state \ref{eq:state}  $N$ times.
    \item Check the frequency of detections. If they are inside the acceptance value, continue to the following step. If not, repeat step 1.
    \item Send the public sequence $Q_2$ for public verification. If the sequence passes the previously established criteria by the parties, continue to the following step. If not, go back to step 1.
    \item  Given the noise model, estimate the minimum compression needed for epsilon secure randomness.
\end{enumerate}



\section{Experiment and Results}

\subsection{Spatial multiplexing scheme}
\begin{figure}
    \centering
    \includegraphics[width= 8.4cm]{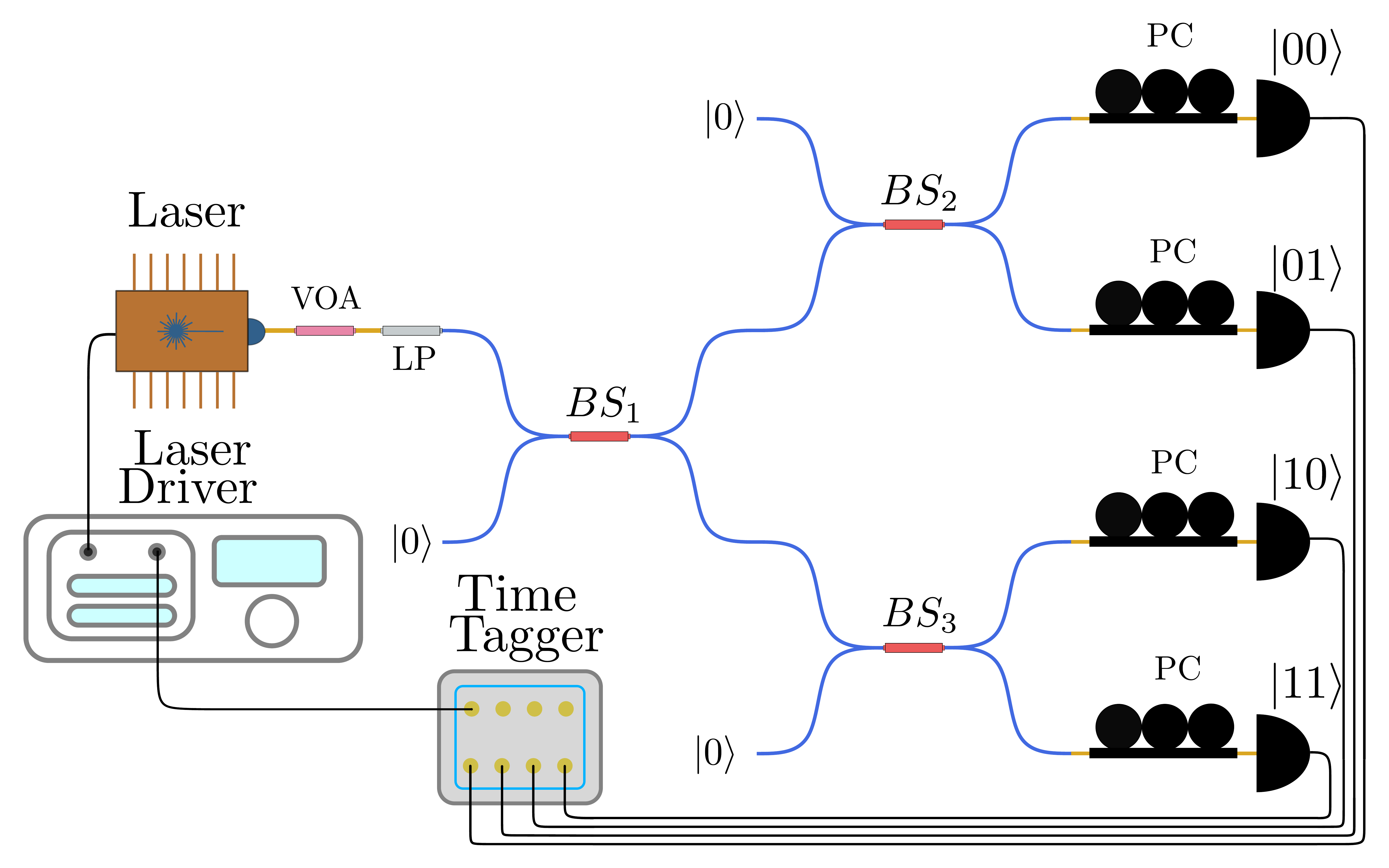}
    \caption{The experimental setup for the $4$-detector scheme involves a pulsed laser attenuated to the Weak Coherent Regime. The laser passes through a Polarization Maintain fiber (blue) and various BS, which results in a four-level superposition state. The Yellow line represents a single-mode fiber attached to a PC controller. The controller is used to control the single-click ratio in each channel. }
    \label{fig:setup_1}
\end{figure}


The experimental setup is illustrated in Fig. \ref{fig:setup_1}. We employed a 1310 nm pulsed-wave laser operating at 3 MHz, attenuated to work in the weak coherent regime \cite{paraiso2021advanced}. The laser beam was directed through an on-fiber linear polarizer (LP) and split into four optical paths using three 50/50 beam splitters (BS). Detection was performed using a Superconducting Nanowire Single Photon Detector (SNSPD) system. Between the beam splitters and the detectors, we introduced a Polarization Controller (PC) to manage the detection rate in each channel by leveraging the polarization sensitivity of the SNSPD. Using the multiplexing capabilities of this setup \cite{buller2009single}, we post-selected single-photon events \cite{jonsson2020photon}, generating the quantum state described by Eq. \ref{eq:state}.

\begin{figure}
\centering
\includegraphics[scale=0.5]{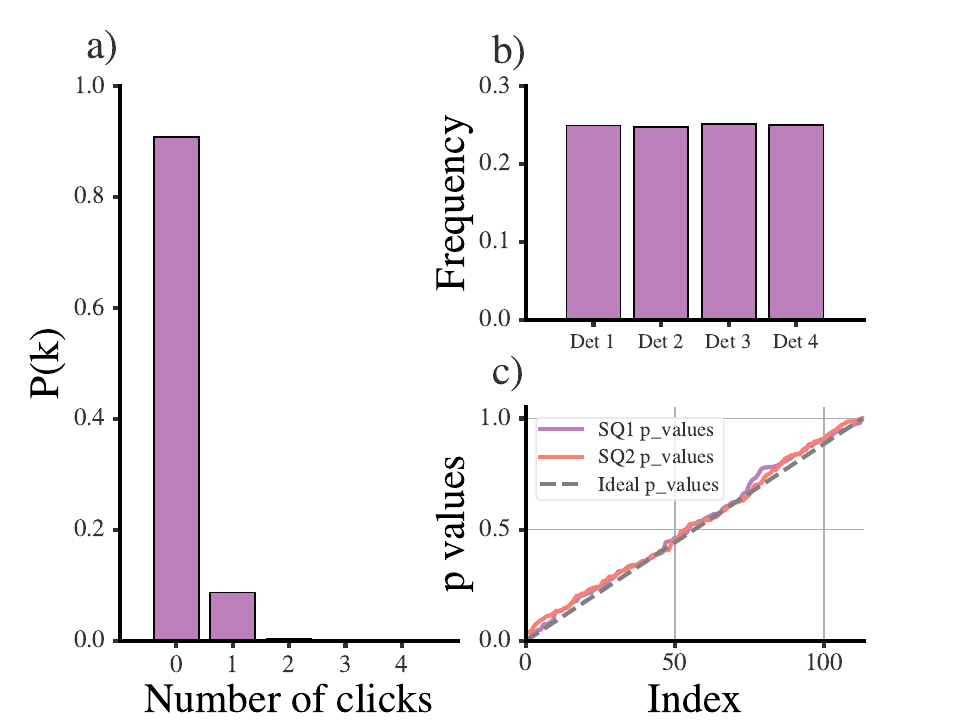}
  \caption{a) Distribution of detected single-photon events across the four detectors.b) Sorted p-values from the Dieharder randomness test suite. Black dashed line represents the ideal expected trend. Colored lines indicate tests for sequences $S_{Q_{1}}$ and $S_{Q_{2}}$, suggesting near-ideal performance of our source.}
  \label{fig:multiplexing_results}
\end{figure}

\begin{figure}
    \centering
    \includegraphics[width= 8.4cm]{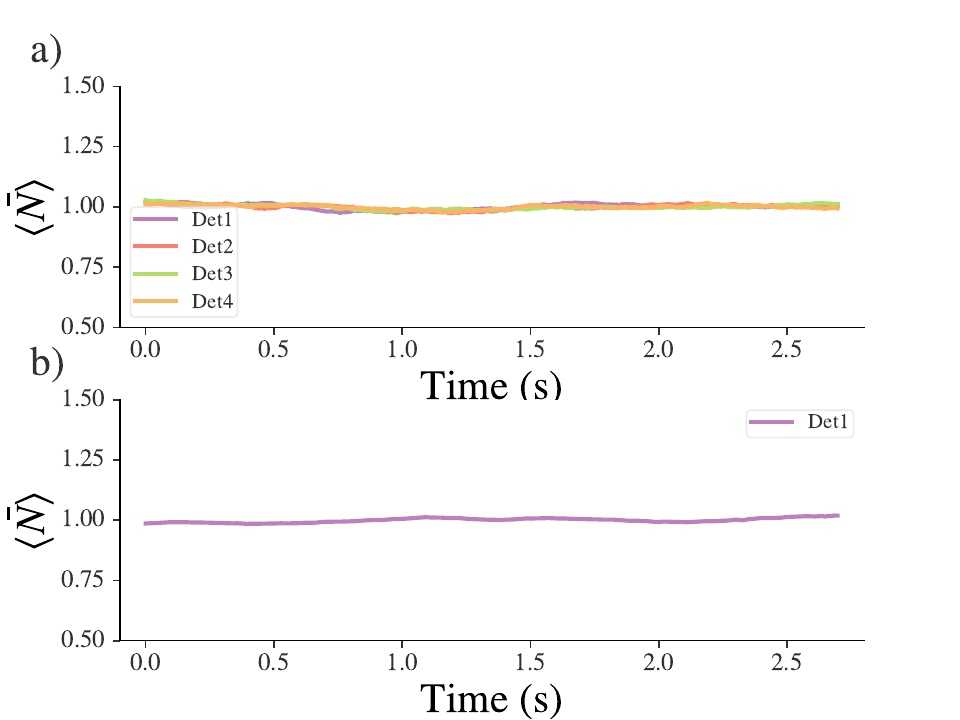}
    \caption{The following figure shows the fluctuation over the mean number of photons as a function of time measured in two different schemes: (a) spatial multiplexing and (b) temporal multiplexing. Each point is calculated by taking the mean value over 100 intervals of $3\cdot10^{-3}$ s. The mean number depicted in the figure is divided by the mean number taken over the entire interval. The fluctuation observed is less than $1.2\%$ for all cases.}
    \label{fig:pol_drift}
\end{figure}

During the measurement, we observed a drift in the detection probabilities at each output channel, attributed to polarization drift in the single-mode fiber (SMF) sections of the setup. To mitigate this issue, we implemented 1-second control intervals and only acquired data when the relative count rates across detectors were constrained within the range of 0.24 to 0.26. This ensured that the entropy distance between sequences, quantified as $|\sqrt{(H(Q_1)-H(Q_2))^2}|$, remained bounded below 0.001. Before initiating data acquisition, we verified the operational conditions of the setup. First, we confirmed that the system operated within the weak coherent regime by measuring the frequency of multi-click events per laser pulse. The results, shown in Fig. \ref{fig:multiplexing_results}a), validated that multi-photon contributions were negligible. Next, we tested the stability of detection probabilities to ensure they remained constant over the measurement period. This step is crucial, as the equivalence of entropy between the sequences $Q_1$ and $Q_2$ holds under the assumptions of our model, provided the probabilities remain invariant during data collection. To evaluate stability, we recorded 2.8 seconds of raw data, divided it into intervals of $\Delta t = 3 \cdot 10^{-3}$ s, and calculated the mean photon count within each interval. These values were normalized by the mean count across all intervals in the dataset. The results, displayed in Fig. \ref{fig:pol_drift}a), show no significant fast fluctuations, confirming the system's stability.

Once operational stability was confirmed, we proceeded with data acquisition, storing a total of 1.25 Gb of photon events at a click rate of 270,000 clicks per second. Fig. \ref{fig:multiplexing_results}b) shows the distribution of single-photon detection rates across the four output channels for the complete dataset. The acquired data were then processed by mapping each detector click to a two-bit sequence, as defined in Eq. \ref{eq:state}. This produced two distinct bitstrings, $Q_1$ and $Q_2$, each containing 1.25 Gb of data.

Our model predicts that, under the given conditions, both sequences $Q_1$ and $Q_2$ should exhibit similar statistical properties. This relationship enables us to use one sequence to verify the statistical randomness of the other. For this purpose, we selected $Q_2$ for verification. The criteria for randomness acceptance can vary depending on the intended application. For instance, cryptographic use cases may demand more restrictive acceptance criteria to ensure robust security. In our implementation, we determined that $Q_2$ would be deemed valid if it passed all relevant Dieharder statistical tests using their standard parameters. After performing the tests, we observed that $Q_2$ successfully passed all assessments, confirming its statistical randomness under the chosen criteria.

To further illustrate the behavior of both sequences, Fig. \ref{fig:multiplexing_results}c) shows the cumulative density function (CDF) of the p-values obtained from the Dieharder tests for both $Q_1$ and $Q_2$. These are compared against the ideal uniform distribution, represented as a straight line. The results demonstrate that both sequences exhibit random behavior, with their CDFs closely aligning with the expected uniform distribution. The methodology of this comparison follows established statistical protocols \cite{pareschi2012statistical,reezwana2022quantum}. It is important to emphasize that this plot is not intended to compare $Q_1$ and $Q_2$. The Dieharder tests evaluate each sequence independently, and the results simply confirm that both sequences exhibit random properties when subjected to the same statistical tests. This reinforces the reliability of the randomness generation process but does not imply equivalence or superiority of one sequence over the other.

In the final step, we performed post-processing to extract epsilon-secure randomness based on a noise model for our system. Noise sources included dark counts in the SNSPDs and multi-photon events that could cause single clicks. The final compression rate was calculated as $R = 0.830$, given $\alpha = 0.304$ and $\epsilon = 2^{-100}$. For details on the noise model and compression methods, refer to the Appendix.


\subsection{Temporal multiplexing scheme}

\begin{figure*}[!htb]
    \centering
    \includegraphics[width= 17 cm]{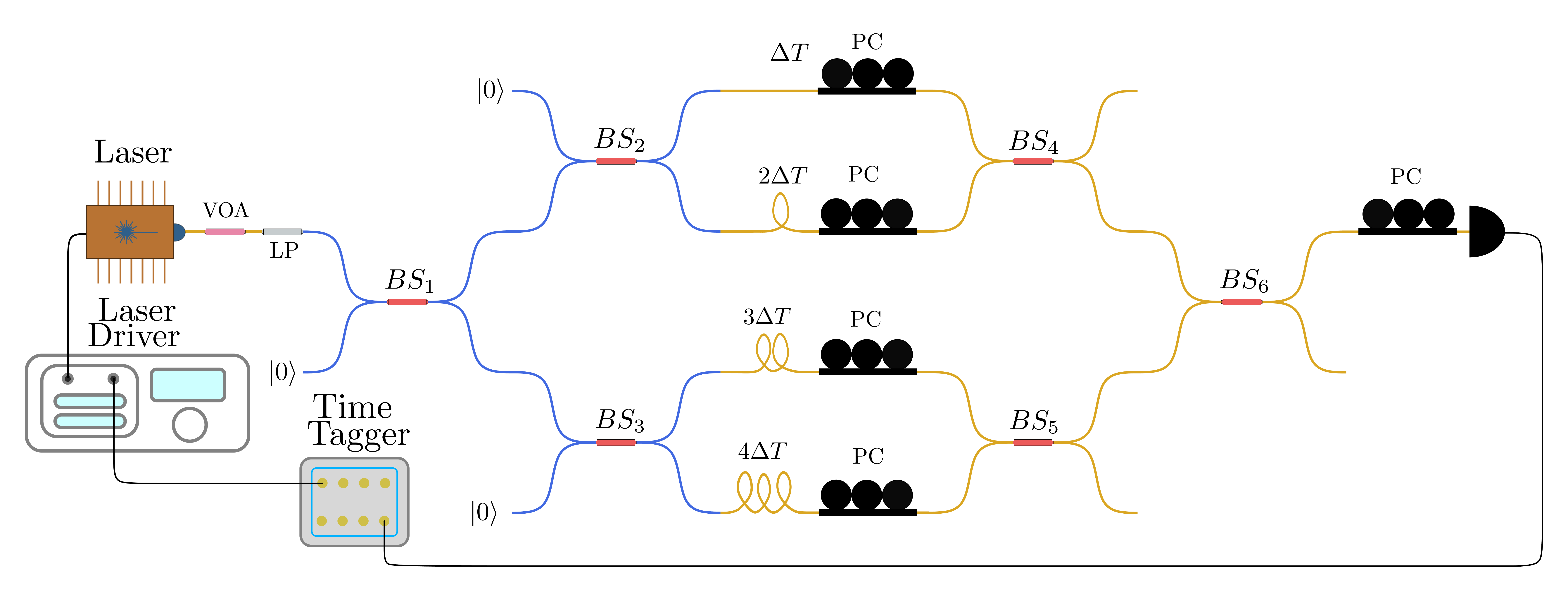}
    \caption{Experimental setup for the 1-detector temporal multiplexing scheme: After passing through beam splitters and a series of fibers with variable lengths to introduce time delays, photons are recombined into separate temporal modes. Polarization controllers ensure a consistent photon click rate across all paths.}
    \label{fig:setup2}
\end{figure*}

To improve the protocol's stability and reduce the number of active elements, we implemented a temporal multiplexing scheme that utilizes a single detector. The experimental setup, shown in Fig. \ref{fig:setup2}, introduces time delays in each path using fibers of varying lengths, ensuring temporal separation between consecutive paths exceeds the detector’s dead time (25 ns). The delayed paths were recombined into a single optical path, with the photon’s path choice mapped onto the time domain. Photon detection was performed using a single SNSPD, while PCs ensured consistent click rates across all temporal modes. Bit encoding was achieved by comparing the photon’s arrival time with the pulse emission time, allowing the generation of the state described in Eq. \ref{eq:state}.

As in the spatial multiplexing scheme, polarization drift caused fluctuations in detection probabilities. To mitigate this, we employed the same balance routine, defining control intervals of 1 second and accepting data only when the relative detection probabilities across temporal modes fell within the range of 0.24 to 0.26. This approach ensured stable detection rates throughout the measurement.

\begin{figure}[h]
\centering
\includegraphics[scale=0.45]{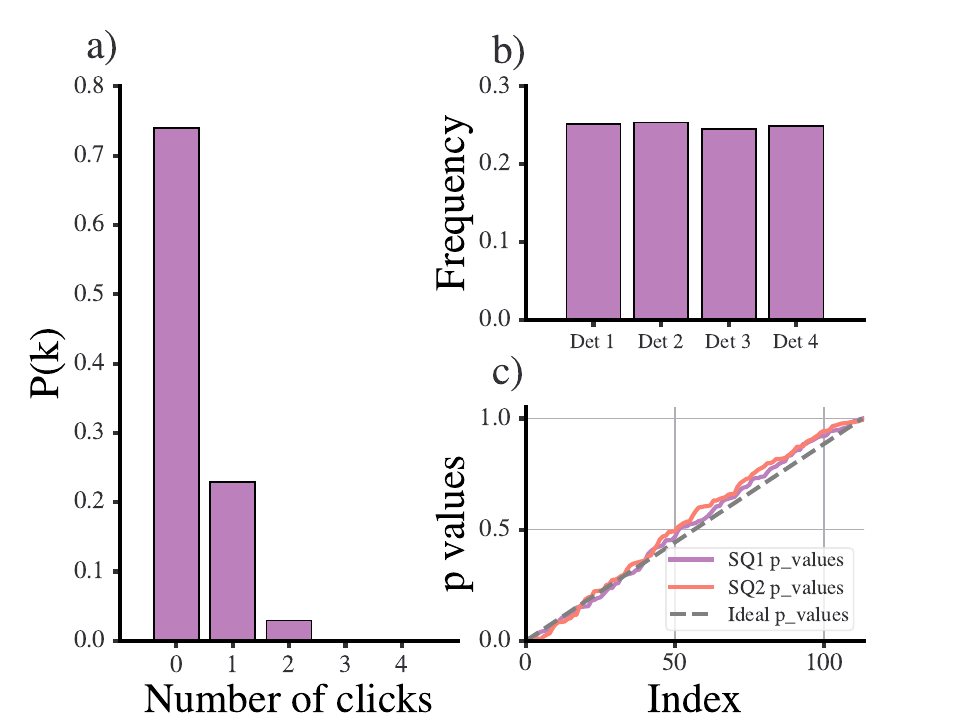}
  \caption{a) Frequency of photon detection events within the designated temporal window, confirming operation in the single-photon regime. b) Sorted p-values from the Dieharder randomness test suite for the temporal multiplexing scheme. }
  \label{fig:temporal_multiplexing_results}
\end{figure}

To confirm proper operation of the setup, we verified that the system was in the weak coherent regime and that the detection probabilities remained stable during the measurement period. Fig. \ref{fig:temporal_multiplexing_results}a) shows the total number of simultaneous clicks, confirming operation in the desired regime. Stability was further validated by analyzing raw data over 2.8 seconds, with the results presented in Fig. \ref{fig:pol_drift}b). Because polarization effects are fully decorrelated across the temporal modes, any fast fluctuations in polarization over this time period would result in noticeable variations in the detection probabilities at the single detector. This would manifest as fast changes in the probability of each output, which would be evident in the plot. The absence of such behavior in Fig. \ref{fig:pol_drift}b) confirms the stability of the system.

Using this configuration, we acquired 1 Gb of photon events at a rate of 590,232 single clicks per second. The proportion of unique clicks across temporal modes is shown in Fig. \ref{fig:temporal_multiplexing_results}b), demonstrating balanced detection rates. The bitstring $Q_2$ was analyzed to assess its randomness. Fig. \ref{fig:temporal_multiplexing_results}b) displays the sorted p-values from the Dieharder statistical tests, confirming that $Q_2$ passed all tests. Additionally, the cumulative density function (CDF) of the p-values was compared to the ideal uniform distribution, and the results aligned closely with the expected behavior, reaffirming the random properties of the sequence.

Finally, post-processing was performed to extract epsilon-secure randomness, accounting for classical noise such as detector dark counts and multi-photon events. Using the parameters $\alpha = 0.574$ and $\epsilon = 2^{-100}$, we achieved a compression rate of $R = 0.868$, reflecting the efficiency of this scheme.

\section{Discussion}
This work presents a simplified protocol based on \cite{jacak2020quantum} that enables public verification of a sequence generated by a QRNG. The protocol was implemented using two experimental schemes: a spatial multiplexing configuration and a temporal multiplexing configuration. Analysis of the generated sequences demonstrated that both exhibit similar statistical properties when evaluated using the Dieharder statistical test. The advantages, limitations, and security considerations of each scheme were examined in detail.

Although the statistical tests confirm the randomness of one sequence, this does not guarantee that the second sequence will pass the tests. However, the verification of one sequence can still be interpreted as a simultaneous test of the entropy source used to generate both sequences, provided the entropy source behaves consistently throughout the generation process. This interpretation assumes that any deviations affecting one sequence would similarly impact the other.

The temporal multiplexing scheme, while currently affected by polarization drift, could be made more robust by using APDs that are not polarization-dependent and attenuators to control the click rates in each path. Under these conditions, changes in detector efficiency would uniformly affect the entire sequence, simplifying performance characterization. Furthermore, if such improvements were implemented, the interpretation of public verification would extend beyond a test of the entropy source to a comprehensive test of the entire device, as both sequences would be measured using identical components.

The protocol could also be adapted to include auditing and verification mechanisms. For example, assigning a third bit to an auditing party, computed as the XOR of the two bits generated after each measurement, would allow the auditor to store an auxiliary sequence. In cases where the private sequence requires reconstruction, the auditor could provide this stored sequence. Such a mechanism has potential applications in scenarios such as lotteries, where the integrity and verifiability of randomness are essential.
\newpage

\bibliographystyle{ieeetr} 
\bibliography{ref.bib}

\appendix
\section{Quantum Randomness Extraction and Noise Mitigation in Photon-Based Quantum Random Number Generation}

The random numbers generated by a QRNG are obtained through a measurement process. During the process, the signal measurement is mixed with noise such as background detection and electronic noises. In cryptography, this noise may be known to Eve, or worse manipulated by her. The main goal of post-processing is to extract quantum randomness and eliminate any contribution of classical noise, thereby erasing all of Eve's knowledge about our sequence.

In our setup, the quantum randomness source is created by preparing a single photon state that passes through three beam splitters. However, for a realistic description of this QRNG, we need to consider that the photon detectors are susceptible to dark counts, and that the efficiencies depend on the number of incoming photons, and that the photon source emits photons according to the Poisson distribution.

To quantify the extraction in our system, we need to determine the probability that the clicks we measure correspond to a single photon event in the presence of inefficient detectors, dark counts, and a Poissonian source given that we measure a single click in our setup.

The photon counting distribution for the case where we send $N$ photons to an array of $n$ detectors given the hypothesis described above is given by \cite{jonsson2020photon}:

\begin{equation}
	\begin{split}
		Pr(k|N;\eta,p_{d},n)  =   \frac{1}{n^{N}}\binom{n}{k}\sum_{l=0}^{k} (-1)^{l}(1-p_{d})^{n-k+l}\times\\
		\times \binom{k}{l}[n-(n-k+l)\eta]^{N}
	\end{split}
\end{equation}

where $k$ corresponds to the number of detectors clicking simultaneously, $p_{d}$ is the probability of dark count in our system, and $\eta$ is the efficiency.

Then, the probability of having $k$ clicks in our system given a probabilistic source $P(N)$ is:

\begin{equation}
p(k)=\sum P(N) Pr(k|n).
\end{equation}

Post-selecting cases with one photon and assuming a Poissonian source, we can calculate the probability of having one click as:

\begin{equation}
\begin{split}
P(k=1)=\exp{-|\alpha|^{2}}\{Pr(k=1|N=0)+\\
|\alpha|^{2} Pr(k=1|N=1)+...\}.
\end{split}
\end{equation}

We can then apply the Bayes rule to calculate the Probability to have $1$ photon given that we detect $1$ click:

\begin{equation}
Pr(N=1|k=1)=Pr(k=1|N=1)\cdot \frac{P(N=1)}{P(k=1)}
\end{equation}

We can identify this probability with the Quantum Bit Error Ratio (QBER) as $Q=1-Pr(N=1|k=1)$.
Next, we can calculate the compression needed as:

\begin{equation}
l_{q}=T\cdot(1-Q)-2\cdot \log_{2}{\Bigg(\frac{1}{\epsilon}\Bigg)},
\end{equation}

Where $T$ is the number of bits before extraction, $l_{q}$ is the length of the bitstring after extraction, and $\epsilon$ represents the security parameter in terms of the statistical distance from perfect randomness.




\end{document}